\documentclass[aps,pra,twocolumn,superscriptaddress]{revtex4}
\usepackage{graphicx}
\usepackage{dcolumn}
\usepackage{color}
\usepackage{bm}
\usepackage{amssymb}
\usepackage{amsmath}
\begin{document}
\title{Coupling field dependent quantum interference effects in a $\Lambda$-system of $^{87}Rb$ atom }
\author{Charu Mishra}
\email{charumishra@rrcat.gov.in}
\affiliation{Laser Physics Applications Section, Raja Ramanna Center for Advanced Technology,Indore 452013, India}
\affiliation{Homi Bhabha National Institute, Mumbai-400094, India}
\author{A. Chakraborty}
\affiliation{Laser Physics Applications Section, Raja Ramanna Center for Advanced Technology,Indore 452013, India}
\affiliation{Homi Bhabha National Institute, Mumbai-400094, India}
\author{Vivek Singh}
\affiliation{Laser Physics Applications Section, Raja Ramanna Center for Advanced Technology,Indore 452013, India}
\author{S. P. Ram}
\affiliation{Laser Physics Applications Section, Raja Ramanna Center for Advanced Technology,Indore 452013, India}
\author{V. B. Tiwari}
\affiliation{Laser Physics Applications Section, Raja Ramanna Center for Advanced Technology,Indore 452013, India}
\affiliation{Homi Bhabha National Institute, Mumbai-400094, India}
\author{S. R. Mishra}
\affiliation{Laser Physics Applications Section, Raja Ramanna Center for Advanced Technology,Indore 452013, India}
\affiliation{Homi Bhabha National Institute, Mumbai-400094, India}

\begin{abstract}
A $\Lambda$-system in $D_2$ line transition of the $^{87}Rb$ atom has been investigated for quantum interference effects for different configurations of coupling field. With a travelling wave coupling field (co-propagating with probe field), the probe beam shows the earlier known electromagnetically induced transparency (EIT) effect at the resonance. With a standing wave coupling field (which is co- and counter-propagating with probe beam), the probe EIT gets transformed into an electromagnetically induced absorption (EIA). A variation in coupling beam power has shown that, at high coupling power, EIA signals have much larger amplitude and slope than those of EIT signals with and without applied magnetic field. These EIA signals can be useful for tight laser frequency locking and optical switching.
\end{abstract}
\maketitle

\section{Introduction}

The quantum interference effects in atomic systems, such as Electromagnetically Induced Transparency (EIT) \cite{Harris:1990} and Electromagnetically Induced Absorption (EIA) \cite{akulshin:1998}, have gained a considerable attention of researchers for interesting physics and upcoming applications. At two photon resonance condition, EIT not only allows the propagation of a probe beam through an otherwise opaque medium, but also produces a large normal dispersion inside the medium. In contrast to EIT, the EIA is constructive quantum interference effect which enhances the absorption of the probe beam along with a large anomalous dispersion inside the medium. These two effects, EIT and EIA, offer potential applications in slow \cite{Kash:1999,Chen:2009,Alotaibi:2015} and fast \cite{Wang:2000} light propagation, high resolution spectroscopy \cite{Motomura:2002, Krishna:2005, Kale:2015}, sensitive magnetometory \cite{Fleischhauer:1994,Lee:1998, Yudin:2010}, precise atomic clocks \cite{Guidry:2017}, optical switching \cite{Clarke:40:2001,Rao:2017}, enhanced non-linear optical processes \cite{Doai:2015,Liang:2017}, precise frequency locking \cite{Bell:2007}, etc.

In the initial studies, the EIT was observed in the three-level systems, namely, lamda-system ($\Lambda$-system) \cite{Li:1995,Yan:2001, Mishina:2011}, Ladder-system \cite{MOSELEY:1995, Bharti:2012} and Vee-system (V-system) \cite{Kang:2014}. In case of EIA, degenerate two level systems \cite{lezama:1999, Goren:2003} and four-level N-system \cite{Goren:2004, Bason:2009} were proved to be the initial milestones. Later on, EIA was also observed in three level $\Lambda$- and Ladder systems by applying an additional coupling field \cite{Whiting:2015,sapam:2012}. The quantum coherence effects, like EIT and EIA, produce signals with extremely narrow spectral-width (sub-natural line-width) which is suitable for tight frequency locking, besides other applications. Also, a complete reversal of the light propagation property in the atomic media, such as transformation of EIT into EIA with an addition of a coupling beam, can be of use  for optical switching. Hence, a detailed study of EIT and EIA phenomena in such atomic systems can be fruitful for fundamental understanding and practical applications. 

In this work, we have studied a $\Lambda$-system in $D_2$ line transition of $^{87}Rb$ atom with a travelling wave coupling field (co-propagating with probe beam) as well as with a standing wave coupling field (co- and counter-propagating coupling waves with probe beam). With the co-propagating travelling wave coupling beam, the probe beam shows an EIT at the resonance. With the standing wave coupling field, which is co- as well as counter-propagating with probe beam, the probe beam transmission shows an electromagnetically induced absorption (EIA). This transformation from EIT to EIA opens the path for medium to be useful for optical switching devices. We have studied this transformation of EIT into EIA by varying the coupling beam power. The EIA signals at higher coupling power have shown dispersion like feature and exhibit steeper slope than the observed EIT signals. The slope of the EIA signals is also higher than the slope of the EIT signals in presence of an applied magnetic field. This steeper slope EIA signal can be utilized for frequency locking of lasers with better stability.

\section{Experimental setup}

\begin{figure}[h]
 \centering
 \includegraphics[width=8.5cm]{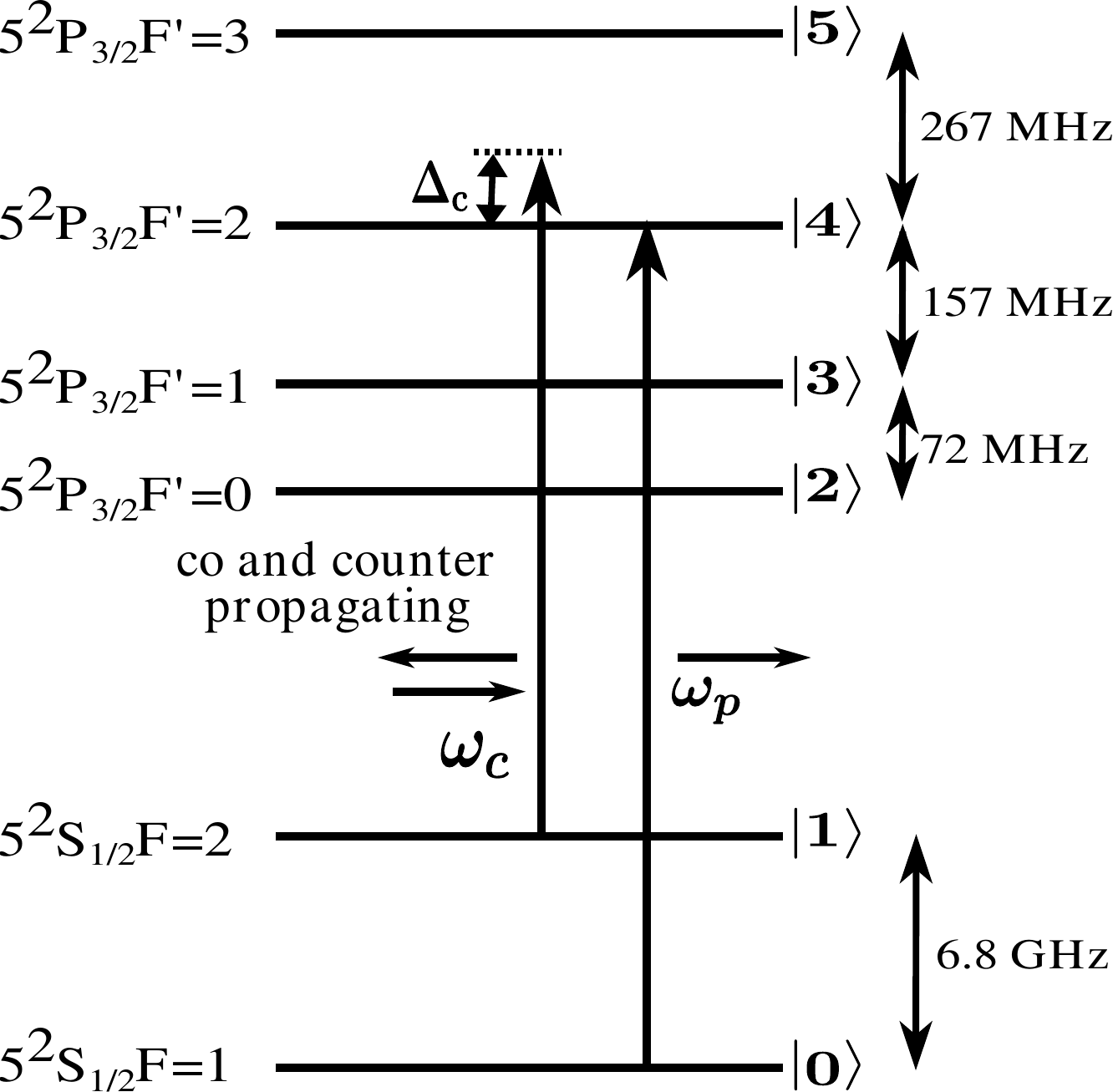}
 \caption{ Relevant energy level diagram of $D_{2}$-line of $^{87}Rb$ showing $\Lambda$-system with a probe beam at frequency $\omega_p$ and two coupling beams at frequency $\omega_c$ with co- and counter-propagating to the probe beam.}
\label{system}
\end{figure} 

\begin{figure}[h]
 \centering
 \includegraphics[width=8.5cm]{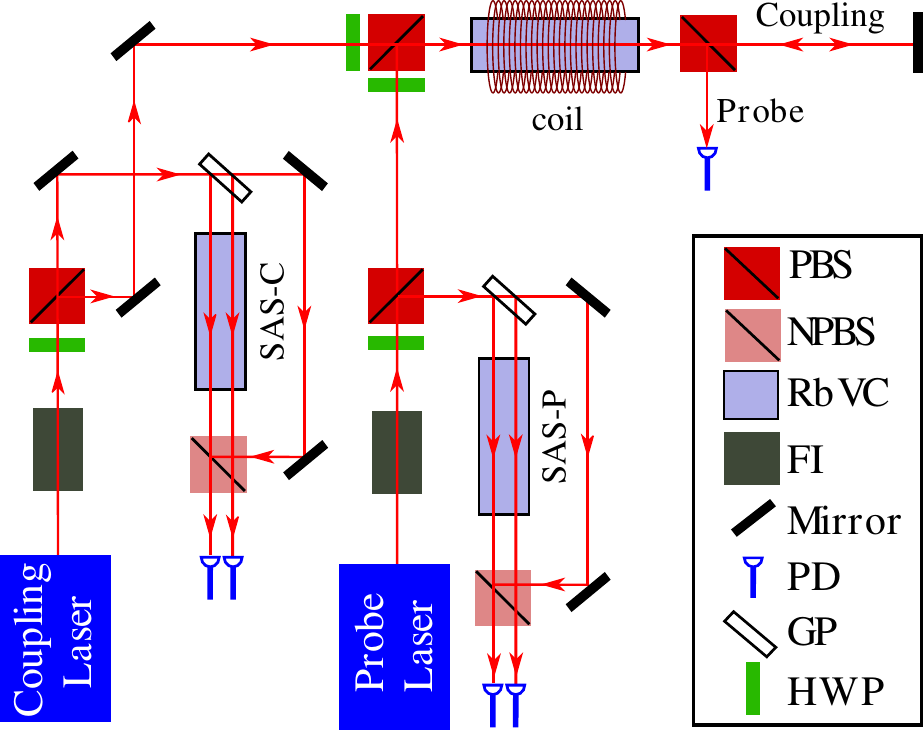}
 \caption{(Color online) Schematic of the experimental setup. PBS: polarising beam splitter; NPBS: non-polarising beam splitter; RbVC: Rubidium vapor cell; FI: Faraday isolator; PD: photodiode; BS: beam splitter; HWP: half-wave plate; BD: beam dump; SAS-P: saturated absorption spectroscopy for probe beam; SAS-C: saturated absorption spectroscopy for coupling beam.}
\label{setup}
\end{figure}

The relevant energy level diagram of $^{87}Rb$ atom is shown in Fig. \ref{system}. A weak (low power) probe beam with frequency $\omega_p$ connects the states $|0 \rangle$ and $|4\rangle$. A strong beam (called coupling beam) at frequency $\omega_c$ couples the states $|1 \rangle$ and $|4\rangle$. The schematic of the experimental setup is shown in Fig. \ref{setup}. Two extended cavity diode lasers, TA-Pro and DL-100 (both from TOPTICA, Germany), operating at wavelength of 780 nm and spectral line-widths less than 1 MHz, were used to derive the probe and coupling beams respectively. The $1/e^{2}$ radii of the probe and coupling beams were 1.36 mm and 2.04 mm, respectively. These beams together formed a $\Lambda$-system by involving transitions ($5^{2}S_{1/2} F=1) \rightarrow  (5^{2}P_{3/2} F'=2) \leftarrow (5^{2}S_{1/2} F=2$). During the experiment, the frequency of the probe beam was kept fixed at the peak of the transition($5^{2}S_{1/2} F=1) \rightarrow (5^{2}P_{3/2} F'=2$) and the frequency of coupling beam was scanned around the transition ($5^{2}S_{1/2} F=2) \rightarrow (5^{2}P_{3/2} F'=2$). These transition peaks were identified using the Doppler free saturated absorption spectroscopy (SAS) technique.

During the experiments, a part of each of the probe and coupling beams was directed to their respective saturated absorption spectroscopy setup (SAS-P/C) to know the frequency/detuning of probe and coupling beams with respect to a hyperfine transition. The probe and coupling beams were passed through a Rb vapor cell of length 50 mm to measure the probe transmission for EIT or EIA observations. The Rb vapor cell contains both the isotopes of Rb, with their natural abundance $\sim$72 \% for $^{85}Rb$ and $\sim$28 \% for $^{87}Rb$. The pressure inside the cell was $\sim 3.6 \times 10^{-7}$ Torr which corresponds to a number density of Rb atoms $\sim 1.2 \times 10^{10} cm^{-3}$. The Rb vapor cell was kept inside two layers of $\mu$-metal sheet to minimize the effect of Earth's magnetic field. Both the beams, with orthogonal linear polarization, were made to co-propagate through the Rb vapor cell at room temperature. The two beams were got separated through a polarizing beam splitter (PBS) kept after the cell. During the experiments, the powers of the probe and coupling beams were controlled using the half-waveplate (HWP) and PBS combinations. For the experiments with the magnetic field, a longitudinal magnetic field was generated using a current carrying coil wrapped over the Rb vapor cell. In case of standing wave coupling field, the transmitted co-propagating coupling beam was got retro-reflected from a mirror to obtain the counter-propagating coupling beam in the vapor cell. For the case of standing wave coupling field, the power indicated $P_c$ denotes only the power of co-propagating coupling beam. To measure the probe transmission, the transmitted probe beam was collected on a photo-diode, after its separation from the coupling beam through a PBS.

\section{Results and Discussion}\label{sec:result}

\begin{figure}[t]
\centering
\includegraphics[width=8.5cm]{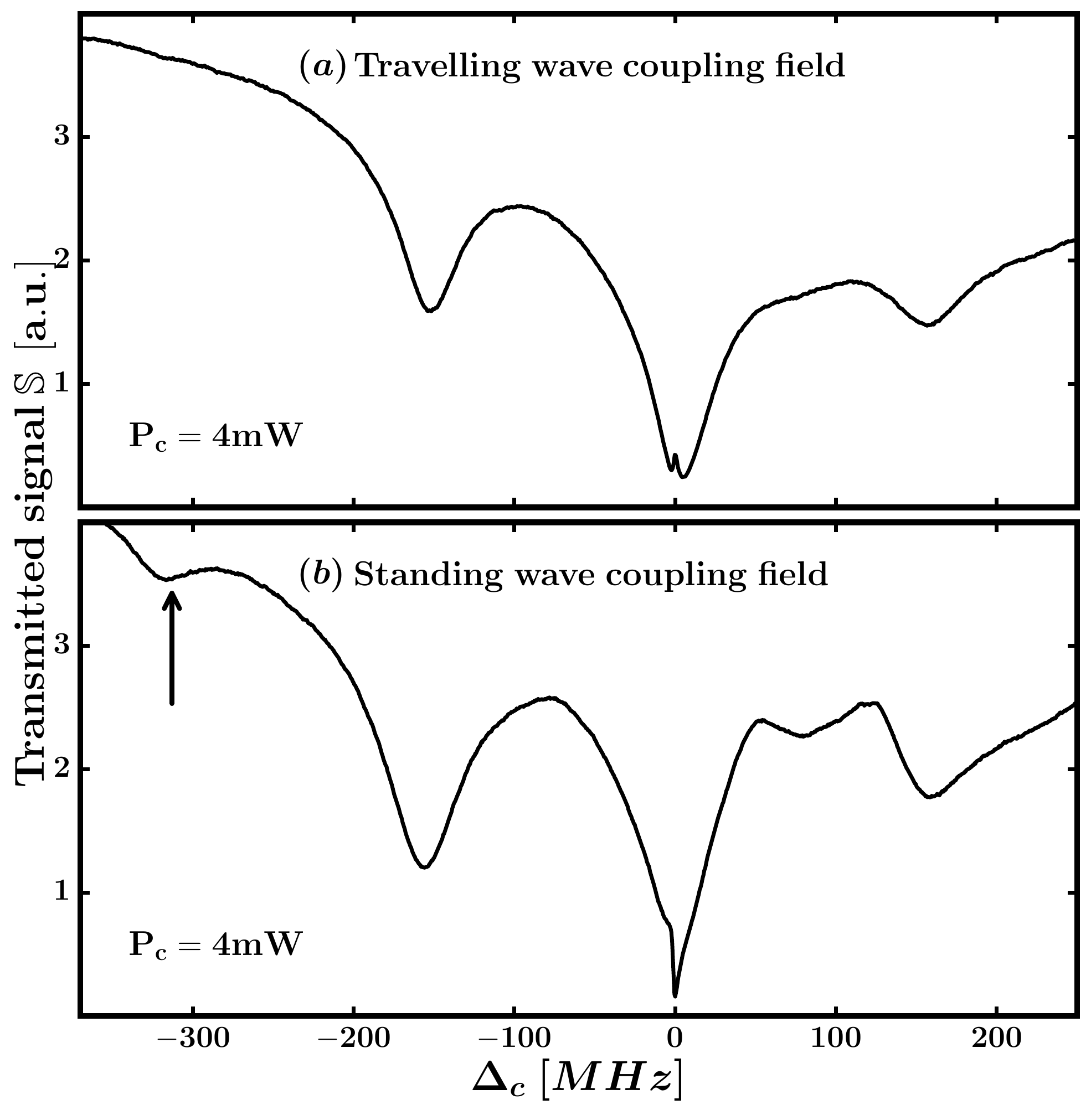}
\caption{ The transmitted probe beam signal as a function of coupling beam detuning. (a) Travelling wave coupling field and (b) standing wave coupling field. Here probe beam power is $\sim$0.08 mW and coupling beam power is $\sim$4 mW}
\label{fig:vsop}
\end{figure}

In the experiments with the co-propagating travelling wave coupling field, the interaction of the moving atoms with the probe beam results in three velocity selective optical pumping absorption dips (VSOP) \cite{Hossain:2011}. In the central absorption dip corresponding to transition between $5^{2}S_{1/2} F=1$ and $5^{2}P_{3/2} F'=2$, an EIT peak emerged with sub-natural linewidth as shown in Fig. \ref{fig:vsop} (a). A detailed study of this $\Lambda$ system with travelling wave coupling field has been reported in an earlier work \cite{Charu:e:2017}. When the retro-reflected coupling beam was passed through the vapor cell to form a standing wave configuration of coupling field inside the atomic medium, the observed transmitted probe beam signal is shown in Fig. \ref{fig:vsop} (b). The presence of retro-reflected coupling beam resulted in increased absorption of probe beam at resonance, indicating the occurrence of EIA instead of EIT.

\begin{figure}[t]
\centering
\includegraphics[width=8.5cm]{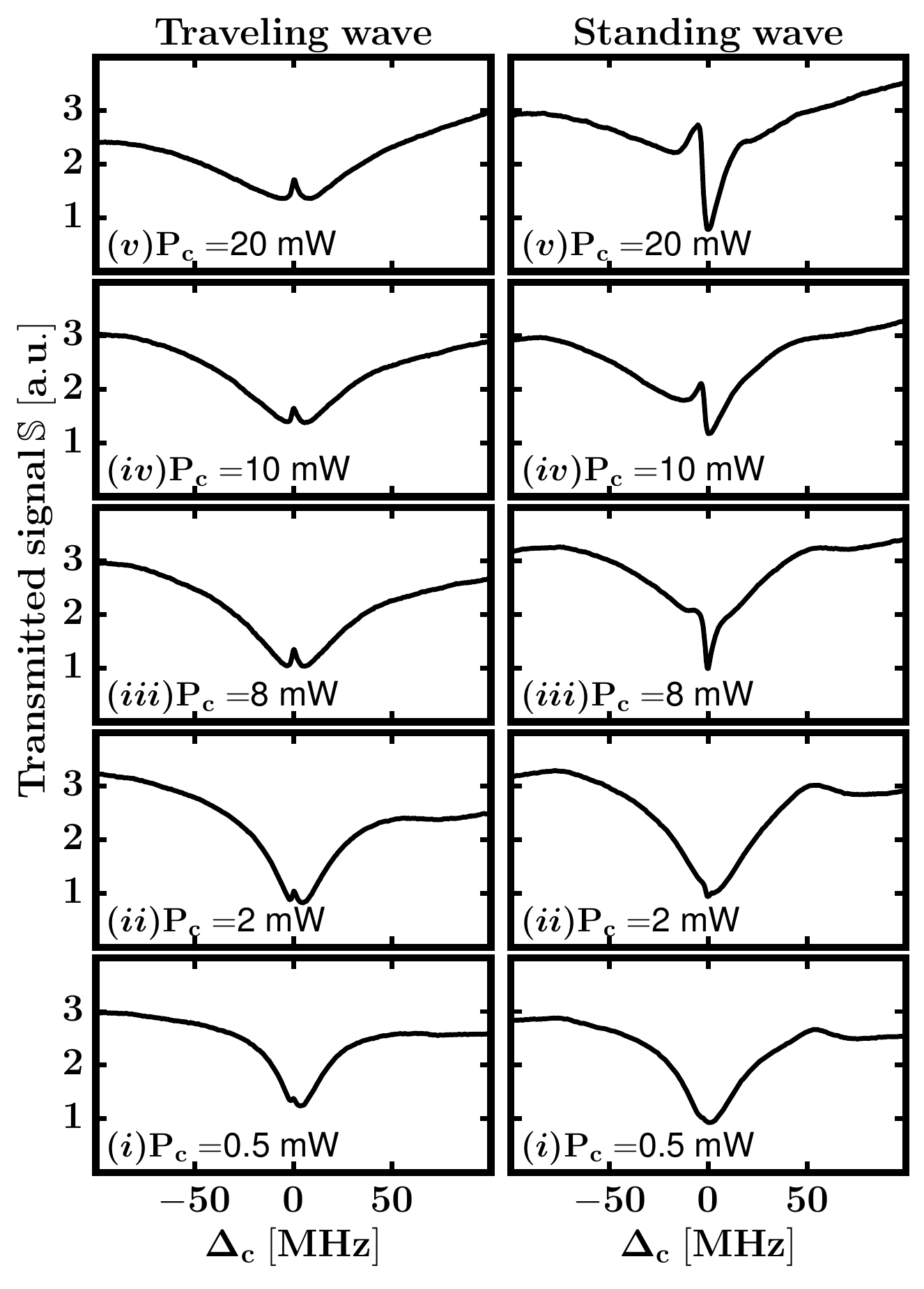}
\caption{ The transmitted probe beam signal as a function of coupling beam detuning for different coupling beam power and fixed probe beam power 0.08 mW. The left and right plots show the spectrum for travelling and standing wave coupling fields respectively. The right plot (for standing wave coupling field) shows power of only forward propagating coupling field.}
\label{fig:spectra}
\end{figure}

The transformation of EIT into EIA, by changing the coupling beam from travelling wave to standing wave, has been reported earlier also \cite{Whiting:2015, Bae:2010}. It may be noted that, in addition to transformation of EIT into EIA, the standing wave coupling field configuration also leads to modification in the VSOP absorption dips. The VSOP dips obtained in presence of the standing wave coupling field were deeper than their travelling wave counterparts due to the participation of a larger number of atoms in the VSOP processes. Apart from this, an additional VSOP absorption dip at coupling beam detuning $\Delta_c \sim -314$ MHz was observed in the standing wave coupling field configuration which is marked by a black arrow in Fig. \ref{fig:vsop} (b). This additional VSOP dip arises due to interaction of a particular group of atoms with the probe beam and only counter-propagating coupling beam.

\begin{figure}[t]
\centering
\includegraphics[width=8.5 cm]{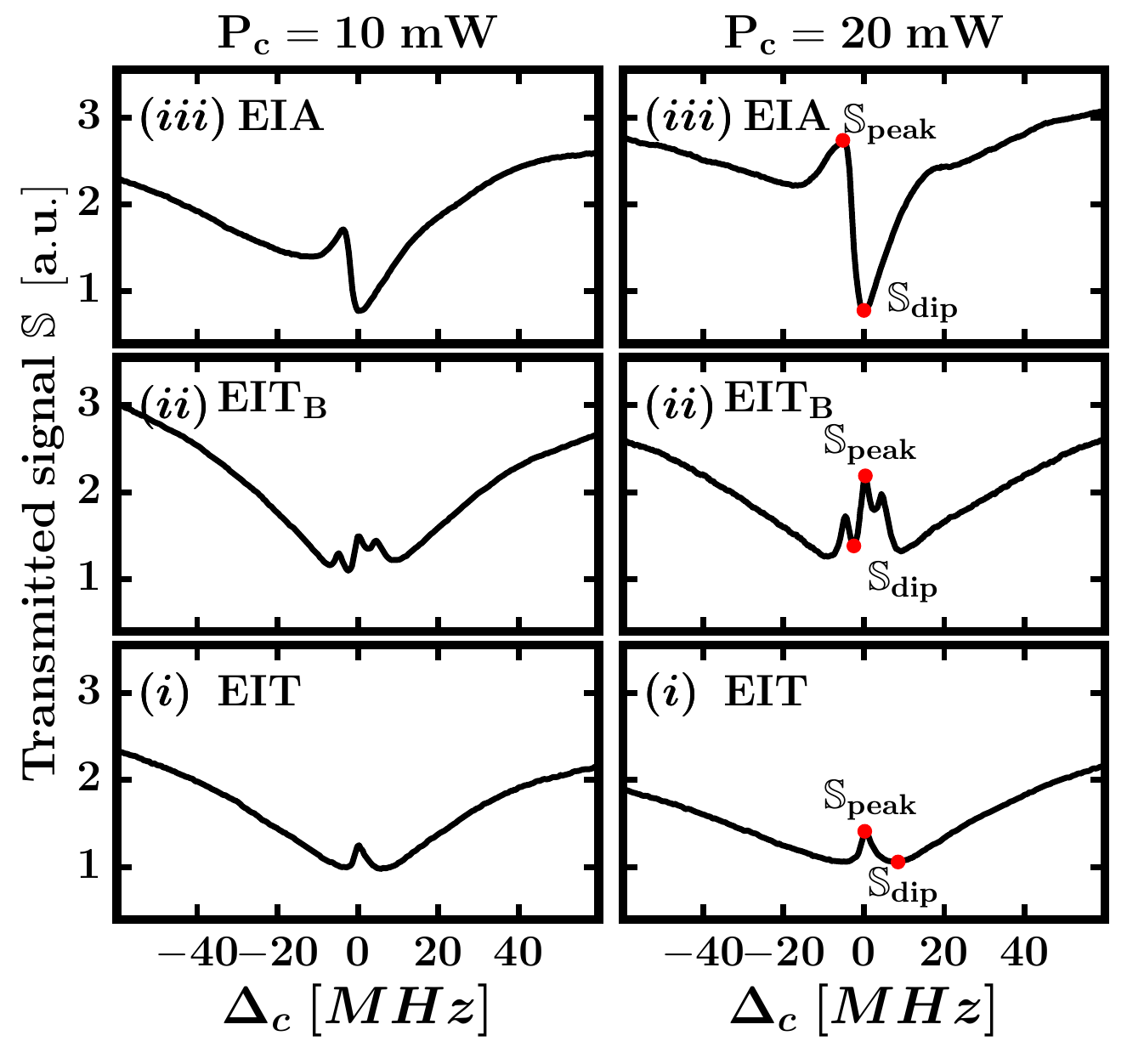}
\caption{ (Color online) Shape of the signals observed in EIT and EIA measurements for two coupling beam powers. Left and right columns show spectrum for $P_c=$ 10 and 20 mW respectively for three different conditions. Plots $(i)$ corresponds to EIT signal in absence of magnetic field, $(ii)$ corresponds to magnetic field assisted EIT signal ($EIT_B$) and $(iii)$ corresponds to EIA signal. Red dots show values used to determine slope of the signal.}
\label{fig:comp}
\end{figure}

Next, we have investigated the EIT and EIA of probe beam by varying the power of the coupling beam. A collection of probe beam transmitted signals for different values of coupling beam powers are shown in Fig. \ref{fig:spectra}. As the coupling beam power was increased, in case of the travelling wave, the amplitude of the EIT signals increased with coupling beam power. Similarly, in case of the standing wave coupling field, the increase in the coupling beam power above $P_c=2$ mW resulted in increase in EIA depth in the transmitted probe signal. It was observed that, for travelling wave coupling power as low as $P_c=0.5$ mW, the EIT signal observed with the travelling wave disappears as retro-reflected coupling beam is introduced. 

The change in the fundamental behavior of quantum coherence in the standing wave coupling field has been dealt previously by several groups \cite{Feldman:1972, Kyrola:1981}. The interaction of an atom with a standing wave coupling field in a $\Lambda$ configuration, has been presented earlier by \citet{Roso:1983}. They have shown that the standing wave coupling field influences the weak probe absorption via three different processes. The first process is known as the population effect, which accounts for the modification in the velocity distribution of populations in different levels via multi-photon processes induced by coupling beam. The second process involves the modification in the spectral response of an atom due to velocity dependent multi-photon interaction and ac-stark shift effects. The last process considers the stationary atoms where the spectral response of these atoms depend upon its position due to spatial variation in intensity in the standing wave field. Later on, results of this dressed state picture were used by several groups to analyze the quantum interference effects in these multi-level atomic systems \cite{Silva:2001, Bae:2010, Chen:61:2014}.

An important feature of this EIA signal is its larger slope and amplitude than EIT signal observed with travelling wave coupling field. In order to establish the supremacy of the EIA signal, a comparison of EIA and EIT signals (in absence and in presence of a magnetic field) was made for different values of power in the coupling field. In one of our earlier studies, it was established that the slope of an EIT signal can be increased by more than two fold by applying an axial magnetic field \cite{Charu:e:2017}. Therefore, EIT in presence of magnetic field was included in this comparison. The results of comparison are shown in Fig. \ref{fig:comp}. 

\begin{figure}[b]
\includegraphics[width=8.5 cm]{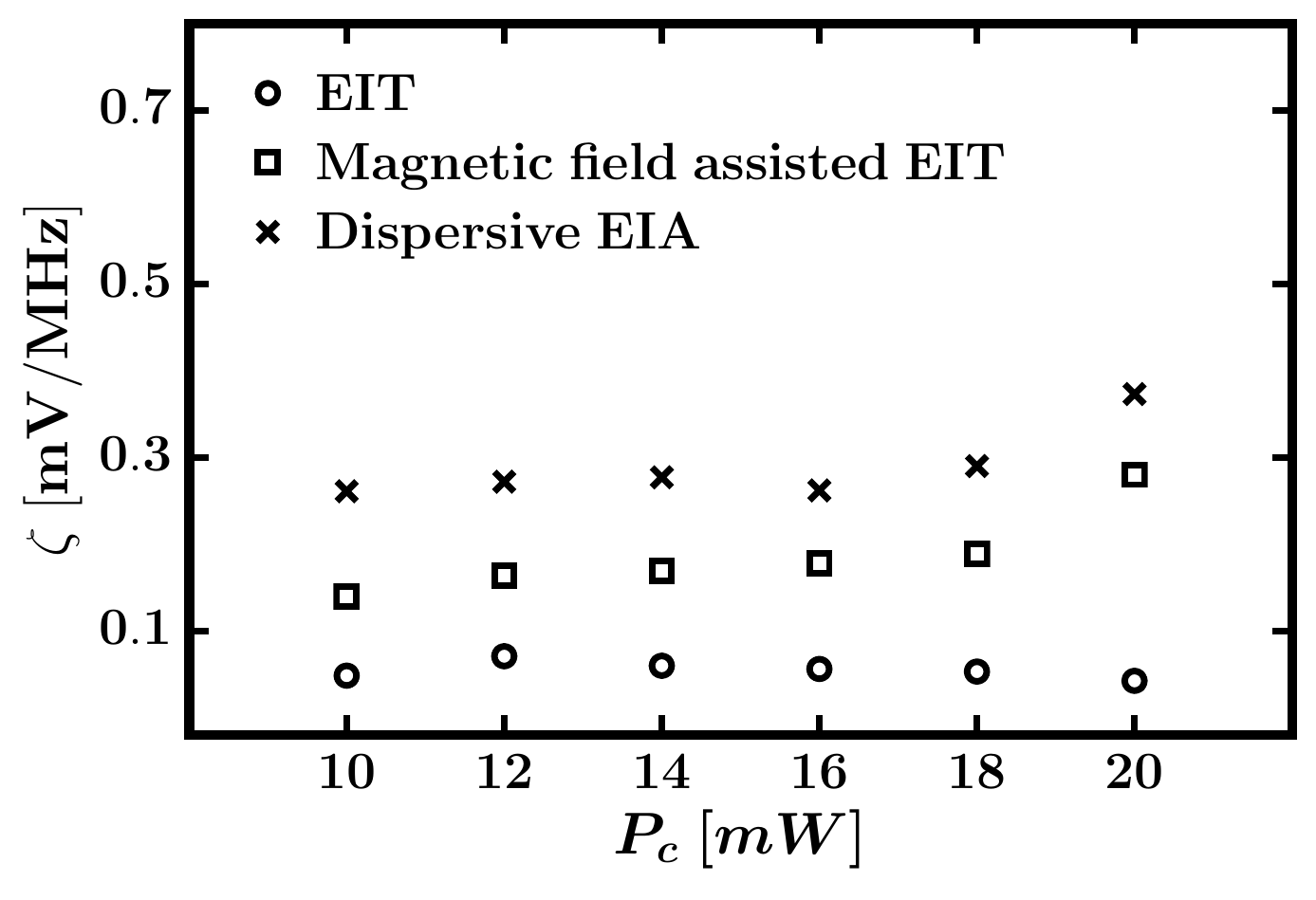}
\caption{ The variation in measured parameter $\zeta$ with coupling beam power for signals obtained in three different cases. Circles, squares and cross represents $\zeta$ values for EIT without magnetic field, EIT with magnetic field and EIA signals, respectively.}
\label{fig:slope}
\end{figure}

As the coupling beam power was increased, the amplitude and slope both were increased. In order to calculate the slope in a signal, we have taken (in Fig. \ref{fig:comp}) the ratio of the difference between the transparency values at peak and dip with the difference in the corresponding detuning values. This is  mathematically defined as, 
\begin{equation}
\zeta=\left\vert\frac{\mathbb{S}^{peak}-\mathbb{S}^{dip}}{\Delta_c^{peak}-\Delta_c^{dip}}\right\vert,
\end{equation} 
where  $\mathbb{S}^{peak/dip}$ is transparency value of peak/dip and $\Delta_c^{peak/dip}$ is corresponding value of coupling beam detuning as shown by red dot in Fig. \ref{fig:comp}.

We note that a higher value of $\zeta$ is preferable for any application like tight frequency locking and optical switching devices, where a reduced relative frequency uncertainty results in higher stability and performance. The values of $\zeta$ for all the three cases have been determined and the results are shown in Fig. \ref{fig:slope}. An unambiguous interpretation can be made using this Fig. that the EIA is better than both the EIT cases. The slope of the EIA signal was significantly higher than that of the EIT signal. The application of these EIA signals in the field of tight laser locking can be promising over generally used EIT signal.

\section{conclusion}

Quantum coherence effects in different configurations have been studied experimentally in a $\Lambda$-system of $D_2$ line transition in $^{87}Rb$ atoms. The conversion of the coupling beam, from travelling to standing wave configuration, leads to conversion of the EIT signal into EIA signal. This conversion may find applications in optical switching. At higher power, the EIA signal has shown much steeper slope than the EIT signal. The higher slopes of the EIA signal compared to the EIT and magnetic field assisted EIT signal is promising for tight frequency locking of lasers.

\section{ACKNOWLEDGMENTS}

We are thankful to A. Chaudhary for his help during experiments. Charu Mishra is grateful for financial support from RRCAT, Indore under HBNI programme.


\end{document}